\documentclass[12pt,letterpaper]{article}

\usepackage{boxedminipage}
\usepackage[pdftex]{graphicx}
\usepackage{amssymb,amsfonts,amsmath}
\usepackage{color}
\usepackage{natbib}
\usepackage{hyperref}
\usepackage{amsthm,mathrsfs}
\usepackage[normalem]{ulem}
\usepackage{graphicx,psfrag,epsf}
\usepackage[affil-it]{authblk}
\usepackage{url} 
\usepackage{xr}
\usepackage{afterpage}
\usepackage{soul} 

\newcommand{\blind}{0}

\newtheorem{proposition}{Proposition}
\newtheorem{lemma}{Lemma}

\newcommand{\mb}{\mathbf}
\newcommand{\mbb}{\mathbb}
\newcommand{\mc}{\mathcal}
\newcommand{\bs}{\boldsymbol}
\newcommand{\trans}{\mathrm{T}}
\newcommand{\indep}{\rotatebox[origin=c]{90}{$\models$}}

\newcommand{\ttne}{\texttt{ne}}
\newcommand{\ttpa}{\texttt{pa}}

\begin{document}

\def\spacingset#1{\renewcommand{\baselinestretch}%
{#1}\small\normalsize} \spacingset{1}


\if0\blind
{
  \title{\bf Bayesian Structure Learning in Multi-layered Genomic Networks}
  \author{Min Jin Ha$^1$, Francesco Stingo$^2$, and Veerabhadran Baladandayuthapani$^3$\\
   \textit{$^1$ Department of Biostatistics, The University of Texas MD Anderson Cancer Center\\
   $^2$ Department of Statistics, Computer Science, Applications ``G. Parenti", The University of Florence\\
    $^3$ Department of Biostatistics, University of Michigan}}
  \maketitle
} \fi


\bigskip
\begin{abstract}
Integrative network modeling of data arising from multiple genomic platforms provides insight into the holistic picture of the interactive system, as well as the flow of information across many disease domains including cancer. The basic data structure consists of a sequence of hierarchically ordered datasets for each individual subject, which facilitates integration of diverse inputs, such as genomic, transcriptomic, and proteomic data. A primary analytical task in such contexts is to model the layered architecture of networks where the vertices can be naturally partitioned into ordered layers, dictated by multiple platforms, and exhibit both undirected and directed relationships. We propose a multi-layered Gaussian graphical model (mlGGM) to investigate conditional independence structures in such multi-level genomic networks in human cancers. We implement a Bayesian node-wise selection (BANS) approach based on variable selection techniques that coherently accounts for the multiple types of dependencies in mlGGM; this flexible strategy exploits edge-specific prior knowledge and selects sparse and interpretable models. Through simulated data generated under various scenarios, we demonstrate that BANS outperforms other existing multivariate regression-based methodologies. Our integrative genomic network analysis for key signaling pathways across multiple cancer types highlights commonalities and differences of p53 integrative networks and epigenetic effects of BRCA2 on p53 and its interaction with T68 phosphorylated CHK2, that may have translational utilities of finding biomarkers and therapeutic targets.
\end{abstract}

\noindent%
\vfill

\newpage
\spacingset{1.45} 
\section{Introduction}
Cancer is a complex disease that is caused by deregulation of several molecular processes and cellular pathways, usually triggered by genetic alterations in specific sets of genes \citep{hanahan2000hallmarks,hanahan2011hallmarks,creixell2012navigating,creixell2015kinome}. Pathway and network analysis aim to gain insight into these underlying interactive mechanisms, and to reduce data involving thousands of altered genes and proteins to a smaller, and more interpretable set of functionally altered processes \citep{pe2011principles}. The Cancer Genome Atlas (TCGA) has generated a rich source of multi-dimensional genomics ('omics for short) data for patients across multiple tumor types and their subtypes (\url{http://cancergenome.nih.gov}). More recently, The Cancer Proteome Atlas (TCPA) (\url{http://bioinformatics.mdanderson.org/main/TCPA:Overview}) has generated a complementary set of reverse phase protein array (RPPA)-based proteomic data across most of these patients' samples, covering major oncogenic signaling pathways \citep{li2013tcpa,akbani2014pan}. Multi-level integration and processing of network information across these modalities is emerging, based on the principle that any biological mechanism is a systematic conflation of multiple molecular events and their interactions \citep{kristensen2014principles}. 

\begin{figure}[h!]
{
\centering
\vspace{-20pt}
\includegraphics[scale=0.2]{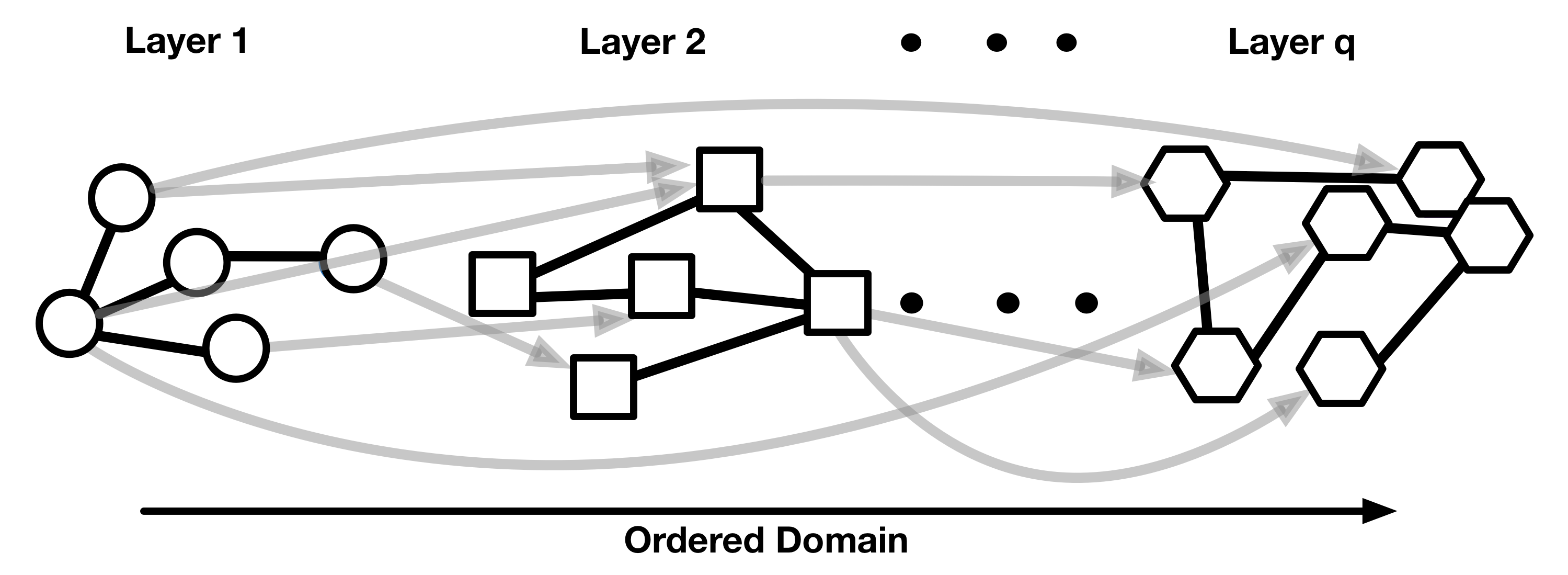}
\vspace{-20pt}
\caption{Example network for $q$ ordered layers.In our application, we consider four layers corresponding to DNA methylation, Copy number aberration, gene expression and protein expression (see Section 7).}
\vspace{-10pt}
\label{fig:net_concept} 
}
\end{figure}

The basic data structure constitutes of a finite sequence of hierarchically-ordered datasets for each individual subject, which facilitates integration of diverse inputs such as genomic, transcriptomic, and proteomic data to investigate the unified regulatory mechanisms underlying the biology of various cancers. Figure~\ref{fig:net_concept} shows a conceptual data structure that allows for the characterization of dependencies within and between the datasets from the ordered layers. Two types of edges characterize the network structure: undirected edges within each layer, and directed edges to distinguish variables in a layer from variables in all the previous layers. 
Most integrative analyses rely on directed relationships between different data platforms based on the biological mechanisms \citep{wang2013ibag}, and the variables observed from each platform constitute a layer. For example, following the central dogma of biology, the first layer could constitute DNA-level data (copy number, methylation), the second layer the transcriptomic (mRNA expression), and the third layer proteomic data. In this context, the directed edges capture cross-platform (e.g., transcriptional, translational) dependencies, and the undirected edges capture the within-platform dependencies.

Modeling each of the layers independently does not account for the hierarchical multi-layered structure of the data, and can only provide information on how networks operate at a static point in time or under a static condition.  A critical next step is to understand a holistic and dynamic picture of the interactive system and the information flow, which can only be achieved by simultaneously modeling multiple layers of data. Our objective is to investigate the conditional dependencies among the variables from multiple layers, while accounting for the order defined by the underlying layered structure of the data. Statistically, this translates to a structural estimation of graphs with a mixture of directed and undirected edges. This objective poses significant methodological and computational challenges, however, to building a single graphical model that includes hierarchical multiple graphs with directed edges for dependencies, constrained by orders between layers, as well as with undirected edges for unconstrained dependencies within layers.

Chain graphs have been used to model the layered architecture of networks, where the vertices can be naturally partitioned into ordered sets that exhibit undirected and directed acyclic relations within and between the sets. The conditional independencies that correspond to missing edges (i.e., Markov properties in the chain graphs) have been studied by \cite{lauritzen1989graphical, frydenberg1990chain,andersson2001alternative}. \cite{drton2006maximum} introduced a point-estimation approach to maximum likelihood estimation given the graph structure, and \cite{drton2008sinful} proposed a constraint-based method to estimate the structure. However, all of the above-mentioned methods for chain graphs are restricted to low-dimensional data with sample sizes larger than the number of vertices. 

In its simplest form (two layers), a chain graph is equivalent to a multiple multivariate regression model. Multiple predictors affect multiple responses that exhibit a correlation structure, and both regression coefficients (for directed edges) and the error precision matrix (for undirected edges) are assumed to be typically sparse. Approaches based on the doubly-penalized joint likelihood have been proposed by \cite{rothman2010sparse,yin2011sparse}. A joint $L_1$ penalty was imposed on both the regression coefficients and the precision matrix, resulting in a penalized likelihood that is bi-convex (but not convex), which implies that the optimization algorithm may be unstable and fail to converge \citep{lee2012simultaneous}. \cite{cai2012covariate,chen2016asymptotically} proposed a two-step approach, in which only regression coefficients are estimated in the first step, and then the error inverse covariance matrix is obtained in the second step, given the estimated regression parameters. To select good initial parameters for this bi-convex problem, \cite{Lin2016} proposed an $L_1$ penalized maximum likelihood estimation with prescreening of variables, and extended this methodology to multi-layered graphs with an arbitrary number of layers. In a Bayesian framework, \cite{bhadra2013joint,SJOS:SJOS12273} proposed a Bayesian model based on the hyper-inverse Wishart prior on the covariance matrix, which assumes that the UGs within each layer are decomposable. The decomposability assumption is generally restrictive, explores a smaller model space, and potentially provides misspecification of the structure for networks in many real applications. 

In this article, we propose a novel approach for a multi-layered Gaussian graphical model (mlGGM) that accommodates data from an arbitrary number of layers. The mlGGM is a special case of the chain graph model, where the random variables that correspond to the vertices are assumed to follow a multivariate Gaussian distribution, and the zero structures of the mean parameter and the precision parameter directly link to the absence of the directed and undirected edges, respectively. We construct a regression-based formulation that converts the mlGGM into a more tractable node-wise multiple regression model. 
We generalize the node-wise multiple regression approach that coherently accounts for conditional independencies in mlGGMs. We jointly select both undirected and directed edges that point to a vertex via Bayesian variable selection priors for each of the regressions, allowing for a much larger graph space than that of decomposable graphs. Moreover, the prior formulation allows for the incorporation of relevant prior knowledge through the edge-specific informative prior, and provides a computationally more efficient procedure to estimate mlGGMs. 

Through simulation studies under various settings of the mlGGMs, we demonstrate the utility of our node-wise regression framework in the structural recovery of the mlGGM, and compare its performance to those of related multivariate regression-based methods. We also numerically evaluate sign consistency (i.e., whether our univariate regression-based formulation correctly finds the signs of the undirected and directed dependencies when we have the known structure). We illustrate the applicability and versatility of mlGGMs to infer integrated genomic networks for multi-omic data, using biological hierarchies among the platforms across multiple cancers. The signed topological structure obtained from our method allows for more refined inference of both cross- and within-platform dependencies, including the inhibition/activation between platforms and positive/negative correlations within platforms. Furthermore, we investigate the commonalities and differences in their multi-layered network structures across cancer types, and show translational utilities of these integrative networks.

The rest of this article is organized as follows. Section 2 provides data structure and backgrounds on mlGGMs. Section 3 presents the joint model and the corresponding prior construction. We introduce our Bayesian node-wise selection (BANS) framework in Section 4. We discuss posterior inference on graphical structure estimation in Section 5. Simulation studies are carried out in Section 6. We demonstrate the utility of our method in multi-layered genomic network studies across cancer types in Section 7. Section 8 concludes the article.
\section{Background}
\subsection{Data structure}
We consider a graphical model over $p$ biological units across multi-omic data, $\bs{Y}=(Y_1,\ldots,Y_p)^\trans \in \mbb{R}^p$. Here $p$ could constitute different platform-specific observations (e.g., genes, proteins, etc.), but for ease of conceptual and technical exposition, we use genes throughout. A graph of $\bs{Y}$ can be denoted by $G=(V,E)$, where $V=\{1,\ldots,p\}$ includes all genes across platforms, and $E$ may contain both directed ($\rightarrow$) and undirected edges ($-$) between the genes. We assume to know {\it a priori} that a partitioning $\mc{T}=\{\tau_k|1\leq k \leq q\}$, $q\leq p$, is a family of pairwise disjoint {\it ordered} layers of $p$ genes. The ordered partitioning $\mc{T}$ implies that any edges between the layers are directed. In other words, each layer $\tau_k$ constitutes genes from a platform, and the layers are ordered according to a biological hierarchy. Formally, the partitioning $\mc{T}$ is called a {\it dependence chain}  if 
\begin{equation*}
k<l \implies v\nrightarrow u \textrm{	} \forall u \in \tau_k, v \in \tau_l.
 \end{equation*}
In other words, when $k<l$, any edges between $\tau_k$ and $\tau_l$ point from a vertex in $\tau_k$ to a vertex in $\tau_l$. For each $v\in V$, let $1\leq t(v)\leq q$ be the index, such that $v\in \tau_{t(v)}$. We assume without loss of generality that the vertices are labeled, such that, 
\begin{equation*}
t(u) < t(v) \implies u<v.
\end{equation*}
%
%
\noindent \underline{Factorization based on biological hierarchies.}
The integrative genomic analyses for multi-platform genomics data can be performed based on coherent biological justifications motivated by their hierarchical dependencies. Following the central dogma of biology, where the epigenetic and DNA level, such as methylation and copy number variation, potentially regulate mRNA expression (transcription regulation), which in turn is known to regulate protein expressions (translation regulation) \citep{morris2017statistical}. For our case study, we integrate four datasets from copy number aberration (CNA), methylation, mRNA expression, and protein expression, and the dependence chain ($\mc{T}$) constitutes four ($q$) disjoint sets that have their own unique order: (CNA, methylation) $<$ mRNA $<$ protein (see Figure~\ref{Fig:prior_ordering})

Notationally, if $u\rightarrow v$, the vertex $u$ is a {\it parent} of $v$. Let $\ttpa_v=\{u\in V: u\rightarrow v\in E\}$ and $\ttpa_A = \cup_{v\in A} \ttpa_v \setminus A$ be the sets of parents of $v$ and a subset $A\subseteq V$, respectively.  We assume the dependence chain allows for factorization of DAGs at the layer level (i.e., the graph in Figure~\ref{fig:net_concept} is a DAG of the dependence chain, $\tau_1,\ldots, \tau_q$). We denote the sub-vector of $\bs{Y}$ corresponding to a subset $A\subset V$ by $\bs{Y}_A$. The joint probability distribution of $\bs{Y}$ can be factorized as
\begin{equation}
P(\bs{Y}) = \prod_{\tau\in\mc{T}} P(\bs{Y}_{\tau}|\bs{Y}_{\ttpa_\tau})
\label{eq:factorization}
\end{equation}
\citep{lauritzen1996graphical}.
Through factorization, the network structure in Figure~\ref{fig:net_concept} can be viewed as $q-1$ two-layered models, such as  $\{\bs{Y}_{\tau_m}:m=1,\ldots, k-1\}$ versus $\bs{Y}_{\tau_k}$ for $k>1$, plus a one-layered UG model for $\bs{Y}_{\tau_1}$. \textcolor{blue}{In the example of using DNA methylation, CNA, mRNA expression and protein expression, the multi-layered networks can be constructed from two UG models for CNA and methylation, and two two-layered models for mRNA and protein expressions.}

To complete our chain graph model, we specify the Markov property. For a dependence chain $\mc{T}$, we define the {\it cumulatives} to be the set $\it{C}_l = \cup_{k\leq l}\tau_k$ for $1\leq l \leq q$.  \cite{andersson2001alternative} proposed a Markov property for chain graphs and set the pairwise Markov property for $G$ as
\begin{eqnarray}
u-v \notin E &\implies& Y_u \indep Y_v | \bs{Y}_{\it{C}_{t(v)}\setminus \{u,v\}} \textrm{ for } t(u) = t(v) \nonumber\\
u\rightarrow v \notin E &\implies & Y_{u} \indep Y_v |\bs{Y}_{\it{C}_{t(v)-1}\setminus \{u\}} \textrm{ for } t(u)<t(v). \label{eq:Markovproperty}
\end{eqnarray}
A missing directed edge between two random variables $Y_u$ and $Y_v$ implies that they are conditionally independent, given all other variables in $\tau_1,\ldots,\tau_{t(v)-1}$, while the conditional set of missing undirected edges is all other variables in $\tau_1,\ldots,\tau_{t(v)}$. 

\subsection{Multi-layered Gaussian graphical models}
In this section, we specify each factor in (\ref{eq:factorization}). We assume that the $p\times 1$ random vector $\bs{Y}$ follows the multivariate Gaussian distribution $N(\bs{0},\bs{\Omega}^{-1})$, with the positive definite precision matrix $\bs{\Omega}$. The mlGGM $G=(V,E)$, following the Markov property stated in (\ref{eq:Markovproperty}), is 
\begin{equation}
\bs{Y} = \mb{B}\bs{Y} + \bs{\epsilon} \textrm{, } \bs{\epsilon}\sim N(\bs{0},\bs{\mc{K}}^{-1}),
\label{eq:1}\end{equation}
where $\mb{B}=(b_{vu})$ is a $p\times p$ matrix with $u\rightarrow v \notin E\Leftrightarrow b_{vu}=0$, and $\bs{\mc{K}} = (\kappa_{vu})$ is a positive definite $p\times p$ matrix with $v-u \notin E\Leftrightarrow \kappa_{uv} = \kappa_{vu} = 0$. Then the precision matrix of $\bs{Y}$ is
\begin{equation}
\mb{\Omega} = (\mb{I}-\mb{B})^\trans\bs{\mc{K}} (\mb{I}-\mb{B}),
\label{eq:2}\end{equation}
where $\mb{I}$ is an identity matrix. $\mb{B}$ is a coefficient matrix for which the zero structures encode the directed edges between layers. The precision matrix of the error $\bs{\epsilon}$, $\bs{\mc{K}}$ is a symmetric matrix, where the nonzero off-diagonal elements represent the undirected edges within a layer after taking out the effects from the directed edges, and the $i$th diagonal element is the inverse variance of $Y_i$.

The mlGGM in (\ref{eq:1}) for a dependence chain $\mc{T}$ with $|\mc{T}|=q>1$ can be expressed as one GGM for $\tau_1$ and $q-1$ two-layered GGMs by the factorization in (\ref{eq:factorization}). For a matrix $\mb{A}=(a_{ij})$, we denote sub-matrices $\mb{A}_{\mc{S}_1,\mc{S}_2}=(a_{ij})_{i\in\mc{S}_1,j\in \mc{S}_2}$ and $\mb{A}_{\mc{S}} = (a_{ij})_{i\in \mc{S},j\in \mc{S}}$. We also denote the transpose of a sub-matrix, $\mb{A}^\trans_{\mc{S}_1,\mc{S}_2}=(\mb{A}_{\mc{S}_1,\mc{S}_2})^\trans $. Under the factorization in equation (\ref{eq:factorization}), the model in (\ref{eq:1}) can be re-expressed as the component-wise conditional distributions: 
\begin{equation}
\bs{Y}_{\tau}|\bs{Y}_{\ttpa_\tau} \sim N(\mb{B}_{\tau,\ttpa_\tau}\bs{Y}_{\ttpa_\tau},\bs{\mc{K}}_\tau^{-1})  \textrm{ for all $\tau\in \mc{T}$}\label{eq:blockrecursive}.
\end{equation}
Because the first layer $\tau_1$ has an empty parent set, it has zero mean in (\ref{eq:blockrecursive}) and is equivalent to the UG. The conditional distribution corresponds to a multivariate multiple regression, where the block of variables $\bs{Y}_\tau$ is regressed on the parents $\bs{Y}_{\ttpa_\tau}$.

There are other formulations of chain graphs based on the Lauritzen-Wermuth-Frydenberg (LWF) Markov property \citep{lauritzen1989graphical, frydenberg1990chain}. In the case of continuous variables with a joint multivariate Gaussian distribution, the Markov property in (\ref{eq:Markovproperty}) is coherent with data generated by the linear system in (\ref{eq:blockrecursive}) \citep{cox1993linear,andersson2001alternative,drton2006maximum}. 
Further details on implication of Markov properties are in Section S1.

\section{Joint model and prior construction}
In this section, we discuss the estimation problem of the multi-layered graph structure that includes both directed and undirected edges, which induces network-based integration of multi-omic data. Under known $\bs{\mc{T}}$, our focus is to estimate the zero-structures of $\mb{B}$ and $\bs{\mc{K}}$ that encode the directed edges and undirected edges between and within layers, respectively. For directed edges, the problem boils down to finding the parents, $\ttpa_v$ for $v\in \tau_k\subseteq V$ from the cumulative $\it{C}_{k-1}$, which is the union of all the preceding layers of $\tau_k$. The model corresponding to the first layer, $\tau_1$, is given by 
\begin{equation}
\bs{Y}_{\tau_1} \sim N(\mb{0},\bs{\mc{K}}_{\tau_1}^{-1}) ,
\label{eq:firstblock}
\end{equation}
where there are no predictors. From the second component to the last component, $\tau_2,\ldots \tau_q$ with $|\mc{T}|=q$, the structure of the model is expressed by the following multivariate multiple regressions: 
\begin{equation}
\bs{Y}_{\tau_k} = \mb{B}_{\tau_k,\it{C}_{k-1}}\bs{Y}_{\it{C}_{k-1}} + \bs{\epsilon}_{\tau_k} \textrm{, and }  \bs{\epsilon}_{\tau_k}\sim N(\mb{0},\bs{\mc{K}}_{\tau_k}^{-1}), \textrm{ }k=2,...,q,
\label{eq:otherblock}
\end{equation}
where $\it{C}_k$ is the cumulative for the $k$th layer. The model for the first layer only involves the parameter for the precision matrix $\bs{\mc{K}}_{\tau_1}$, which is a traditional model for UGs \citep{lauritzen1996graphical}. The parameters of interest are the coefficient matrices $\{\mb{B}_{\tau_k,\it{C}_{k-1}}:k=2,\ldots,q\}$, which encode the directed edges across the layers, and $\{\bs{\mc{K}}_{\tau_k}:k=1,\ldots,q\}$, which encode the conditional dependencies within the layers, after adjusting for the effects from the directed edges. Fitting the set of regression models can be performed independently and in parallel, assuming priors over the parameters that are independent across the regressions. 
\noindent \underline{Component-wise regressions.} In this component-wise regression framework, we can use the Inverse Wishart prior for  $\bs{\mc{K}}_{\tau_k}^{-1}$ and the independent normal priors for elements of $\mb{B}_{\tau_k,\it{C}_{k-1}}$: for all $k=1,\ldots, q$,
 \begin{equation}
 \begin{aligned}
 \bs{\mc{K}}_{\tau_k}^{-1} &\sim Inverse \textrm{ } Wishart_{|\tau_k|}(\delta_{\tau_k},\lambda_{\tau_k}\mb{I}_{|\tau_k|}),\\
 b_{vw}  &\sim N(0,c_{vw}^2/\kappa_{vv}) \textrm{ for all $v\in \tau_k$ and $w\in \it{C}_{k-1}$},
 \end{aligned}
 \label{eq:prior1}
 \end{equation}
 where $\lambda_{\tau_k}>0$  and $\delta_{\tau_k}>0$ for all $k=1,\ldots,q $, and $c_{vw}$ are constants. The prior of each regression coefficient $b_{vw}$ is dependent on the precision $\kappa_{vv}$ of the variable $Y_v$ ($v$th diagonal element of $\bs{\mc{K}}$). 
 
 In essence, we are able to divide the estimation problem of the mlGGM in (\ref{eq:1}) into smaller multivariate regression problems by specifying the priors that are independent across $\tau\in \mc{T}$ in (\ref{eq:prior1}). The Wishart distribution is the global conjugate priors for the precision matrices and, equivalently, the covariance matrix follows the Inverse Wishart distribution. However, the number of parameters greatly increases as the total number of variables from multiple layers increases. For estimating the two-layered GGM for the $k$th component, $\tau_k$, we have $|\tau_k| |C_{k-1}| + |\tau_k|(|\tau_k|+1)/2$ number of parameters for the directed edges, undirected edges, and the variances of the variables in $\tau_k$. The number of parameters is 3775, even for $|\tau_k|=50$ and $|C_{k-1}|=50$. Moreover, the number of parameters becomes larger when we handle downstream layers, due to the increasing number of preceding layers. \cite{bhadra2013joint} and \cite{SJOS:SJOS12273} proposed a joint selection of the nonzero elements of $\mb{B}_{\tau_k,\it{C}_{k-1}}$ and $\bs{\mc{K}}_{\tau_k}$, where the undirected relations encoded in $\bs{\mc{K}}_{\tau_k}$ correspond to the limited space of decomposable graphs using the hyper-inverse Wishart prior \citep{dawid1993hyper}. Moreover, the method selects entire columns of the coefficient matrix $\mb{B}_{\tau_k,\it{C}_{k-1}}$, and fails to select single elements of this matrix (i.e., a variable in an upper layer is either connected to all variables in a lower layer or to none), and the same precision matrix $\bs{\mc{K}}_{\tau_k}$ informs the selection of the coefficient matrix $\mb{B}_{\tau_k,\it{C}_{k-1}}$. 
Assumptions on both $\mb{B}_{\tau_k,\it{C}_{k-1}}$ and $\bs{\mc{K}}_{\tau_k}$, while improving computational efficiency due to conjugate formulation and allowing for the exact calculation of the marginal likelihood of the graph, impose the artificial restriction on both undirected and directed structures, and may potentially result in misspecification of the networks structures.
 In particular, the space of decomposable graphs corresponding to $\bs{\mc{K}}_{\tau_k}$ is increasingly sparse with the increasing size of $\tau_k$. For example, the percentages of graphs that are decomposable decrease as 95\%, 80\%, 55\%, 29\% and 12\% for $|\tau_k|=$ 4, 5, 6, 7, and 8, respectively \citep{armstrong2005bayesian}. 
\section{Bayesian node-wise selection (BANS) framework}
To circumvent the above-mentioned challenges, we develop a model selection procedure that allows for more general graph space than the procedures proposed by \cite{bhadra2013joint,SJOS:SJOS12273}, as well as greater computational efficiency than the joint model formulation. We propose a Bayesian node-wise selection (BANS) method to jointly estimate undirected edges and directed edges of mlGGMs, using a node-wise regression framework. For a node, the Bayesian variable selection approach simultaneously finds its neighbors and parents, which are connected by undirected and directed edges, respectively. 
\subsection{Working model}
For all $v\in V$, $\mc{C}_v$ and $\mc{P}_v$ are defined by the set of all other vertices in the same layer as $v$,  $\tau_{t(v)}\setminus \{v\}$ and all the preceding vertices, $C_{t(v)-1}$, respectively. We consider multivariate regression between $\tau_{t(v)}$ as responses and $\mc{P}_v$ as predictors. Note that the sets in $\{\mc{P}_v:v\in \tau\}$ are all the same, and we consider $\mc{P}_v=\mc{P}_\tau$ for $v\in \tau$. We re-express models (\ref{eq:firstblock}) and (\ref{eq:otherblock}) to node-wise regressions:
\begin{equation*}
Y_v = \bs{Y}_{\mc{P}_v}^\trans \mb{b}_v +\epsilon_v \text{ for all } v\in V ,
\label{eq:orgmodel}
\end{equation*}
where $\bs{\epsilon} = (\epsilon_1,\ldots,\epsilon_p)^\trans \sim N(\bs{0},\bs{\mc{K}}^{-1})$ and $\mb{b}_v$ is a $|\mc{P}_v|\times 1$ vector of $\{b_{vu}:u\in \mc{P}_v\}$. 
For the well-known conclusion that gives the relation between the concentration matrix of the multivariate normal distribution and the regression coefficients \citep{anderson1984multivariate}, we have the following lemma.
\begin{lemma}
For all $v\in V$, $\epsilon_v = \bs{\epsilon}_{\mc{C}_v}^\trans\bs{\alpha}_v + e_v$, where $\bs{\alpha}_v = -\bs{\mc{K}}_{\mc{C}_v,v}/\kappa_{vv}$ and $e_v\sim N(0,1/\kappa_{vv})$ is independent of $\bs{\epsilon}_{V\setminus \{v\}}$.
\end{lemma}
\noindent 

The undirected edges encoded in the zero structures of the precision matrix $\bs{\mc{K}}$ can be found by the zero structures of the regression coefficients obtained from regressing each residual $\epsilon_v$ to the residuals corresponding to the other vertices in the same layer. 
\begin{proposition}
For all $v\in V$, $Y_v = \bs{Y}_{\mc{P}_v}^\trans \mb{b}_v + \bs{Y}_{\mc{C}_v}^\trans\bs{\alpha}_v  -\bs{Y}_{\mc{P}_v}^\trans\mb{B}^\trans_{\mc{C}_v,\mc{P}_v} \bs{\alpha}_v + e_v$, where $\bs{\alpha}_v = -\bs{\mc{K}}_{\mc{C}_v,v}/\kappa_{vv}$ and $e_v\sim N(0,1/\kappa_{vv})$ is independent of $\bs{\epsilon}_{V\setminus \{v\}}$.
\label{lemma2}
\end{proposition}
\noindent The proof of Proposition \ref{lemma2} is provided in Section S2. From Proposition \ref{lemma2}, we coherently convert the estimation problem of mlGGM in (\ref{eq:1}) to that of node-wise regressions. For a given node, both undirected edges between vertices in the same layer and directed edges toward the vertex can be jointly uncovered by model selection for a regression model. Our working model can be re-expressed as follows: for all $v\in V$,
\begin{equation}
Y_v = \sum_{i\in \mc{P}_v}b_{vi}Y_i + \sum_{j\in \mc{C}_v} \alpha_{vj}Y_j + \sum_{j\in \mc{C}_v} \alpha_{vj} \sum_{i\in \mc{P}_v} -b_{ji}Y_i + e_v,
\label{eq:working}
\end{equation}
where $e_v\sim N(0,1/\kappa_{vv})$. Our model (\ref{eq:working}) for a vertex $v\in V$ involves regression coefficients for the directed edges from the vertices in the preceding layers ($\mc{P}_v\rightarrow v$), undirected edges from the vertices in the same layer ($\mc{C}_v-v$), and directed edges from the vertices in the preceding layers to the vertices in the same layer other than $v$ ($\mc{P}_v \rightarrow \mc{C}_v$). 

\begin{figure}[h!]
{
\centering
\vspace{-5pt}
\includegraphics[scale=0.3]{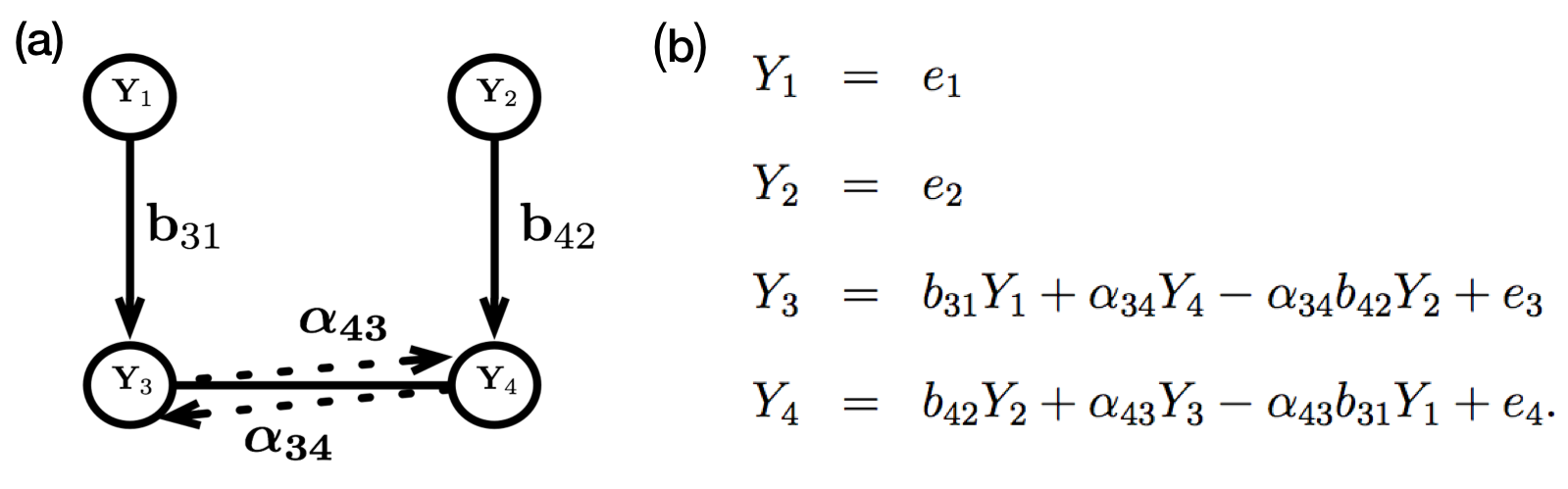}
\vspace{-10pt}
\caption{An example of a chain graph with layers $\{1,2\}$, $\{3,4\}$. For example, the first layer $\{1,2\}$ and the second layer $\{3,4\}$ represent gene and protein expressions, respectively. (a) Solid lines represent the data generating structure and dotted lines display how we parameterize undirected edges in our neighborhood selection framework. (b) Our working model.}\label{fig1} 
\vspace{-5pt}

}
\end{figure}

\noindent \underline{A simple illustrative example}.  Consider the linear model corresponding to the chain graph with two layers for genes $\{Y_1,Y_2\}$ and proteins $\{Y_3,Y_4\}$ described with solid lines in Figure~\ref{fig1} (a): $Y_1 = \epsilon_1$, $Y_2 =  \epsilon_2$, $Y_3 = b_{31}Y_1 + \epsilon_3$, and $Y_4 = b_{42}Y_2 + \epsilon_4$, where $\epsilon_1$ and $\epsilon_2$ and $(\epsilon_3,\epsilon_4)$ are mutually independent, and $\epsilon_3$ and $\epsilon_4$ (the residuals of proteins after taking out effects from genes) have bivariate normal distribution with an arbitrary covariance matrix. The undirected edge between proteins $Y_3$ and $Y_4$ is estimated by the regression coefficients, $\alpha_{43}$ and $\alpha_{34}$, from the two regressions for responses $Y_3$ and $Y_4$ in our working model with true nonzero regression coefficients, as shown in Figure~\ref{fig1} (b).
For example, for the protein $Y_3$, our working model includes effects for the parent gene $Y_1$, the neighbor protein $Y_4$, and the indirect effect of the neighbor's parent gene $Y_2$. Therefore, when we select directed and undirected edges for a biological unit $v$, the indirect effects of its neighbor's parents are adjusted. 

The next step is to deduce the prior distribution of $\{\bs{\alpha}_v|v\in V\}$ from the joint prior on the precision matrix $\bs{\mc{K}}$. The priors in (\ref{eq:prior1}) are re-expressed as follows: for all $\tau\in \mc{T}$,
 \begin{equation}
 \begin{aligned}
 \alpha_{vw}|\kappa_{vv} &\sim N(0,1/(\lambda_{\tau}\kappa_{vv})) \textrm{ for all $v\in \tau$ and $w\in \mc{C}_v$},\\
 \kappa_{vv}& \sim Gamma(\frac{\delta_{\tau}+|\tau|-1}{2},\frac{\lambda_{\tau}}{2}) \textrm{ for all $v\in \tau$},\\
  b_{vw}  &\sim N(0,c_{vw}^2/\kappa_{vv}) \textrm{ for all $v\in \tau$ and $w\in \mc{P}_v$},
 \end{aligned}
 \label{eq:prior_ne}
 \end{equation}
where the Wishart prior on $\bs{\mc{K}}_\tau$ translates to normal and gamma priors on $\bs{\alpha}_v$ and $\kappa_{vv}$ for $v\in \tau$.

\subsection{Model selection priors}
Our goal is to infer the dependence structure of the underlying chain graph of the data $\mb{Y}$. Using our framework, the undirected and directed edges of the chain graph can be selected using zero restrictions on the regression parameters, $\mb{B}$ and $\bs{\alpha}$. Based on the priors in (\ref{eq:prior_ne}), model selection is achieved through a mixture prior on the regression coefficients: for all $v\in \tau$, $w\in \mc{P}_v$, $u\in \mc{C}_v$, and $\tau\in \mc{T}$,
\begin{equation}
 \begin{aligned}
\textrm{Directed edges: } &&b_{vw} | \gamma_{vw},\kappa_{vv} \sim \gamma_{vw}N(0,c_{vw}^2/\kappa_{vv}) + (1-\gamma_{vw})\delta_0,\\
\textrm{Undirected edges: }&&\alpha_{vu}|\eta_{vu},\kappa_{vv} \sim \eta_{vu}N\left(0,1/\left(\lambda_\tau\kappa_{vv}\right)\right) + (1-\eta_{vu})\delta_0,\\
\textrm{Variance parameters: }&&\kappa_{vv} \sim Gamma(\frac{\delta_{\tau}+|\tau|-1}{2},\frac{\lambda_{\tau}}{2}),
\end{aligned}
\label{eq:finalprior}
\end{equation}
where $\delta_0$ is the Dirac delta function and $c_{vw}$, $\lambda_\tau$, and $\delta_\tau$ are fixed hyperparameters. We complete the formulation of our model by specifying the prior on $\gamma_{vw}$ and $\eta_{vw}$:
\begin{equation*}
P(\gamma_{vw} =1) = p_{vw} \textrm{ and } P(\eta_{vw}=1) = q_{vw},
\end{equation*}
where $p_{vw}$ and $q_{vw}$ are fixed hyperparameters. The binary indicators, $\gamma_{vw}$ and $\eta_{vw}$, are latent variables that encode the directed structure between layers and the undirected structure within a layer. If $\gamma_{vw}=1$ , the arrow from $w$ to $v$ $(w\rightarrow v)$ is included in the graph, and $\gamma_{vw}=0$ otherwise. If $\eta_{vw}=1$, the undirected edge between $v$ and $w$ $(v-w)$ is present in the graph. 

\section{Likelihood and posterior inference}
For each vertex, we implement the Bayesian procedure to select the parents from the preceding layers and neighbors from the same layers as the vertex. Let $\mb{Y}=(\mb{y}_1,\ldots \mb{y}_p)$ and $\mb{Y}_A$ for a set $A\subset V$ be a $n\times p$ data matrix and $n\times |A|$ sub-matrix of $\mb{Y}$ for which the columns correspond to the set of vertices $A$. From Proposition~\ref{lemma2}, we have the following node-wise likelihood function: 
\begin{eqnarray}
\scriptsize
\begin{aligned}
L= &\prod_{\tau}\prod_{v\in \tau} p(\mb{y}_v|\mb{Y}_{\mc{C}_v\cup\mc{P}_v},\bs{\alpha}_v,\kappa_{vv},\mb{B}_{\tau,\mc{P}_v})\nonumber \\
 =&\prod_{\tau}\prod_{v\in \tau} \left(\frac{\kappa_{vv}}{2\pi}\right)^{n/2} \exp\left\{-\frac{\kappa_{vv}}{2}\left(\mb{y}_v - \mb{Y}_{\mc{P}_v}\mb{b}_v - \mb{Y}_{\mc{C}_v}\bs{\alpha}_v  +\mb{Y}_{\mc{P}_v}\mb{B}^\trans_{\mc{C}_v,\mc{P}_v} \bs{\alpha}_v\right)^\trans \left(\mb{y}_v - \mb{Y}_{\mc{P}_v}\mb{b}_v - \mb{Y}_{\mc{C}_v}\bs{\alpha}_v  +\mb{Y}_{\mc{P}_v}\mb{B}^\trans_{\mc{C}_v,\mc{P}_v} \bs{\alpha}_v\right)\right\}, 
\end{aligned}
\label{eq:likelihood}
\end{eqnarray}
\normalsize
where $\bs{\alpha}_v=-\bs{\mc{K}}_{\mc{C}_v,v}/\kappa_{vv}$. We construct the likelihood $L$ by pooling conditional densities within each of the two-layer models. The inference using the likelihood $L$ is not equivalent to the joint likelihood from our original model in (\ref{eq:firstblock}) and (\ref{eq:otherblock}), because the resulting local distributions are likely to be {\it inconsistent} in that there is no joint distribution $p(\bs{Y}_\tau|\bs{Y}_{\mc{P}_\tau})$ from each of the local distributions $p(Y_v|\bs{Y}_{\mc{C}_v\cup\mc{P}_\tau})$ \citep{heckerman2000dependency}. The results are asymmetry of the structure, magnitudes, and signs inferred from $\{\bs{\alpha}_v:v\in V\}$ between $\alpha_{43}$ and $\alpha_{34}$ in Figure~\ref{fig1}. The neighborhood selection approach \citep{meinshausen2006high} for UGs requires a symmetrization step after estimating the regression coefficients for all node-wise regressions. Similarly, in our model framework, the set of undirected edges can be defined by $\{v-w: \eta_{vw}=1 \textrm{ and } \eta_{wv}=1\}$ or $\{v-w:\eta_{vw}=1 \textrm{ or } \eta_{wv}=1\}$. For more accurate inference on the graphical structure, instead of using the post-hoc symmetrization, we implement MCMC sampling (BANS) to estimate the posterior distributions with the symmetric constraint $\eta_{vw} = \eta_{wv}$ and showed better accuracy in the structural learning than post-hoc symmetrization obtained from BANS-parallel (Section S7.5 in the Supplementary Materials). However, considering the gain in computation efficiency by using the node-wise parallelizable procedure, BANS-parallel is scalable and useful for high-dimensional problems (Section S7.5 in the Supplementary Materials).

\subsection{MCMC Sampling}
Our neighborhood selection approach for the chain graph model enables us to estimate the chain graph via a vertex-wise variable selection framework. Now we consider estimating the undirected edges and directed edges toward a vertex $v$. Since the parameter spaces for the binary indicators $\bs{\eta}$ and $\bs{\gamma}$ are enormous, computing the explicit posterior probabilities for all possible subsets poses computational challenges. Instead, we use a stochastic search variable selection (SSVS) to generate a Gibbs sequence for each vertex $v$ \citep{george1993variable}. Our sampling scheme consists of two parts for updating undirected edges and directed edges. The symmetric constraints of the UG structure can be incorporated when sampling $\bs{\eta}$ by assuming $\eta_{vw}=\eta_{wv}$ for $v\neq w$. The MCMC algorithm, which is described in detail in Section S3 can be summarized as follows. For a vertex $v\in \tau \subseteq V$ at iteration $t$:\\
1. Undirected edges
\begin{itemize}
\item[1.1] Set $\tilde{\mb{y}}_v =\mb{y}_v-\mb{Y}_{\mc{P}_v}\mb{b}_v $ 
and $\mb{X}_v = \mb{Y}_{\mc{C}_v}-\mb{Y}_{\mc{P}_v}\mb{B}^\trans_{\mc{C}_v,\mc{P}_v}$.
\item[1.2] Update $\left\{\bs{\eta}_k:k\in \tau \right\}$ and set $\ttne_v^{(t)}=\{w\in \tau: \eta_{vw}\neq 0\}$.
\item[1.3] Update $\left\{\bs{\alpha}_k:k\in \{v\} \cup \ttne^{(t)}_v\right\}$, and $\left\{\kappa_{kk}:k\in \{v\}\cup \ttne^{(t)}_v\right\}$.
\end{itemize}
2. Directed edges
\begin{itemize}
\item[2.1] Set $\tilde{\mb{y}}_v = \mb{y}_v - \mb{Y}_{\mc{C}_v}\bs{\alpha}_v$ and $\mb{X}_\tau=\begin{pmatrix}\mb{Y}_{\mc{P}_v}&-\alpha_{vu_1}\mb{Y}_{\mc{P}_v}&-\alpha_{vu_2}\mb{Y}_{\mc{P}_v} & \ldots\end{pmatrix}$, where $u_1, u_2,\ldots$ are vertices in $\ttne^{(t)}_v$.
\item[2.2] Update $\{\bs{\gamma}_v: v\in \tau\}$, $\mb{B}_{\tau,\mc{P}_v}$, and $\kappa_{vv}$.
\end{itemize}
Sampling parameters that correspond to undirected edges are not independent among the vertices in the same layer, due to the imposed symmetric constraints. Thus, our MCMC sampling is performed for each layer. Within an MCMC sampling, steps 1 and 2 are repeated for all vertices in a layer. The mlGGM estimation using this MCMC sampling method is called BANS. We also implemented node-wise sampling scheme, called BANS-parallel in Section S6. 

\subsection{Graphical structure estimation}
The posterior samples of the model parameters for undirected and directed edges obtained from our MCMC methods are used to perform Bayesian inference. The MCMC samples explore the distribution of possible graphs, with each graph leading to a different topology based on the model parameters. The maximum a posteriori (MAP) estimate represents the mode of the posterior distribution of possible graphs. This approach is not feasible, however, since the space of possible graphs is large and the most likely graph may still appear only in a very small proportion of MCMC samples. An alternative and practical solution is to select the edges marginally by using all of the MCMC samples and averaging the presence/absence of each edge over the MCMC samples \citep{hoeting1999bayesian}.

We propose a false discovery rate (FDR)-based determination of significant networks. Our MCMC methods are applied to each of the layers as responses, with all the preceding layers as predictors. For each layer, suppose we have $M$ posterior samples of the corresponding parameter set. From $M$ MCMC iterations for all layers, we estimate the posterior marginal probability of edge inclusion for each edge $g_{vw}$ as the proportion of MCMC iterations after the burn-in in which the edge $v-w$ for $t(v)=t(w)$ or $w\rightarrow v$ for $t(w)<t(v)$ was selected in the graph. The values $1-g_{vw}$ can be considered as Bayesian $q$-values, or estimates of the local FDR \citep{storey2003statistical,newton2004detecting}, if the $vw$th edge is called a discovery. Given a desired FDR level $\alpha\in (0,1)$, we call the set of edges $\mc{X}_{\phi_{\alpha}} =\{(v,w):g_{vw}>\phi_\alpha\}$ discoveries. The significance threshold $\phi_\alpha$ can be determined based on the approach of \cite{baladandayuthapani2014bayesian}. We first sort all $\{g_{vw}\}$ in descending order to yield $\{g_{(t)}\}$. Then we set $\phi_\alpha=g_{(\xi)}$, where $\xi = \max\{k|k^{-1}\sum_{j=1}^k (1-g_{(j)})<\alpha\}$. We expect that only 100$\alpha$\% of the discovered edges $\mc{X}_{\phi_{\alpha}}$ are false positives. An alternative approach is to select the set of edges that appear with marginal posterior probability of inclusion greater than $\phi_\alpha = 0.5$ \citep{barbieri2004optimal}. This rule results in an expected FDR for $\phi_\alpha$
\begin{equation*}
FDR = \frac{\sum_{(v,w)}\left(1-g_{vw}\right) I(g_{vw}>\phi_\alpha)}{\sum_{(v,w)}I(g_{vw}>\phi_\alpha)},
\end{equation*}
where $I$ is the indicator function. 

\section{Simulations}
The aim of our simulation study is to compare BANS with other joint estimation approaches under various simulation settings that generate high- and low-dimensional data under the mlGGM in (\ref{eq:1}) (Section 6.2.1). We also numerically evaluate the sign consistency of the estimated partial correlations for undirected edges within layers and the estimated coefficients $b_{vw}$ for directed edges between layers using our structured sampling approach (Section 6.2.2). We perform sensitivity analysis of our algorithm to priors (Section S4, Supplementary Materials) and check the convergence (Section S5, Supplementary Materials).

\subsection{Data generation with random chain graphs}
To generate random chain graphs, we assume that the UGs within layers follow the Erd\"{o}s and R\'{e}nyi (ER) model \citep{erdos1960evolution}. A graph that follows the ER model is constructed by randomly connecting the vertices. We assume that each undirected edge within a layer is included in the chain graph with probability $p_E$ independent from all other edges. We also assume that each directed edge between two consecutive layers is independently linked with probability $p_E/2$. Thus, the vertex $i$ in the graph is almost surely connected to $(|\tau_{t(i)}|-1)p_E$ undirected edges and $|\tau_{t(i)-1}-1|p_E/2$ directed edges. We generate an adjacency matrix $A=(A_{ij})_{p\times p}$ that represents a random chain graph on $V=\{1,\ldots, p\}$ with $A_{ij}=A_{ji}=1$ for $i-j$, and $A_{ij}=1$ and $A_{ji}=0$ for $i\rightarrow j$. We use the random chain graph generation procedure as follows: (1) assign $p$ vertices to the dependence chain $\mc{T} = \{\tau_k|k=1,\ldots,q\}$ so that the sizes of the layers are mostly the same; (2) set independent realizations of $\text{Bernoulli} (p_E)$ in the lower triangular elements of the sub-matrix corresponding to each layer, then symmetrize it; and (3) set independent realizations of $\text{Bernoulli}(p_E/2)$ in the off-block diagonal elements between two consecutive layers. Given a randomly generated chain graph $G$ with the dependence chain $\mc{T}$, the observed data are simulated by the mlGGM in equation (\ref{eq:1}), after setting the intensities of the nonzero elements of $\mb{B}$ and $\bs{\mc{K}}$. The nonzero elements of $\mb{B}$ and $\bs{\mc{K}}$ are randomly sampled from $(-1.5,-0.5) \cup (0.5,1.5)$. To guarantee the positive definiteness of $\bs{\mc{K}}$, their diagonal elements are filled by column-wise sums of absolute values, plus a small constant. Then we draw a random sample $\bs{Y}$ of size $n$ from the distribution $N(\mb{0},\bs{\Omega}^{-1})$ with $\bs{\Omega}$ in equation (\ref{eq:2}). In the additional simulation setting that emulates the DNA Damage response network estimated for 309 TCGA lung squamous cell carcinoma (LUSC) samples (Section S7.1), we used the parameter estimates $\hat{\mb{B}}$ and $\hat{\bs{\mc{K}}}$ obtained from the real data application to generate simulation datasets.

\subsection{Performance evaluation}
\subsubsection{Graphical structure estimation}
We compare the performance of our BANS method against those of multivariate regression-based methods in learning the topologies of chain graphs under various simulation settings of mlGGMs according to the simulation parameters, the number of variables ($p$), sample size ($n$), number of layers ($q$), and degree of sparsity ($p_E$). We consider the following multivariate regression-based methods for two-layer models: multivariate regression with covariance estimation (MRCE) \citep{rothman2010sparse}, and covariate-adjusted precision matrix estimation (CAPME) \citep{cai2012covariate}, which are available in R packages \texttt{MRCE} and \texttt{capme}, respectively. We apply glasso \citep{friedman2008sparse} for MRCE and CLIME \citep{cai2011constrained} for CAPME to estimate the UG for the first layer, as MRCE and CAPME are not applicable to UGs. 

We investigate the performance of our method under different simulation settings by varying 
$(p,n,q,p_E)$. Different measures of structural difference can be used to assess the performance of our method. We define TP, TN, FP, and FN as the total number of true positive, true negative, false positive, and false negative edges, respectively. For goodness of estimation, we use sensitivity, specificity, and Matthew's correlation coefficient (MCC):
\begin{eqnarray*}
\text{Specificity} & = &\frac{\text{TN}}{\text{TN} + \text{FP}} \text{,   }\text{Sensitivity} =\frac{\text{TP}}{\text{TP} + \text{FN}} \text{,   and }\\\
\text{MCC} &=& \frac{(\text{TP}\times \text{TN})-(\text{FP}\times \text{FN})}{\{(\text{TP}+\text{FP})(\text{TP+\text{FN}})(\text{TN}+\text{FP})(\text{TN}+\text{FN})\}^{1/2}}.
\end{eqnarray*}
MCC ranges from -1 (total disagreement) to 1 (perfect classification), with a larger value corresponding to a better fit.

\begin{table}[h!]
   \caption{Simulation results: model selection performances as measured by sensitivity, specificity, Matthew's correlation coefficient (MCC), the number of edges detected, partial area under the curve (pAUC) at specificity=0.8 and area under the curve (AUC), for four different simulation scenarios, based on 50 replications. Numbers in parentheses are the simulation standard errors.} 
\scriptsize
    \begin{tabular}{cccccccc}
    \hline
     Setting & Method & Sensitivity & Specificity & MCC & Number & pAUC$(0.8)$$^*$ & AUC\\
     $(p,n,q, p_E)$ & & & & &of discoveries & &  \\
    \hline
    (20,200,6,0.3) & BANS & 0.99 (0.042) &0.99 (0.005) &  0.9 (0.041) & 14.3 (0.974) &1 (0.015) &1 (0.003)  \\
       $\#$ true edges= 12 & MRCE & 0.99 (0.039) &0.81 (0.046) &0.46 (0.062) &45.64 (8.312) & 0.79 (0.048)&0.96 (0.015)\\
                           & CAPME &1 (0.016) & 0.74 (0.056) &0.39 (0.049) & 58.84 (9.960)&0.74 (0.055)&0.95 (0.009) \\
                           \hline
    (100,200,6,0.03) & BANS & 0.94 (0.034) & 1 (0.001) &0.87 (0.034) & 49.62 (1.947)&1 (0.001)& 1 (0)\\
   $\#$ true edges= 43     & MRCE & 1 (0.008) & 0.9 (0.021) & 0.27 (0.029) & 529.08 (100.605)&0.81 (0.015)& 0.96 (0.003) \\
			      & CAPME & 1 (0.005) & 0.79 (0.028) &0.18 (0.014) & 1087.76 (137.793) &0.85 (0.006)&0.97 (0.002)\\     
			      \hline
(100,200,10,0.03) & BANS & 0.97 (0.021) & 1 (0.001) &0.83 (0.031) &34.66 (1.825) &1 (0.001) &1 (0)  \\
 $\#$ true edges= 25			      & MRCE & 1 (0.008) & 0.93 (0.013) &0.25 (0.025) & 367.02 (64.609) &0.86 (0.008) &0.97 (0.002)\\
			      & CAPME & 1 (0.008) & 0.85 (0.021) & 0.17 (0.014) &776.6 (105.138)& 0.88 (0.011)&0.98 (0.002)\\ 
        \hline
 (200,100,10,0.03) & BANS  & 0.74 (0.024) & 1 (0.002) & 0.83 (0.015)  & 91 (3.536)& 0.99 (0.002) & 1 (0)\\
  $\#$ true edges=116 & MRCE &0.47 (0.050) &0.98 (0.002) &0.22 (0.017) & 469.04 (40.141) &0.87 (0.009)&0.93 (0.013)  \\
 & CAPME & 0.48 (0.029) & 0.94 (0.009) & 0.13 (0.009) & 1327.98 (174.406) &0.79 (0.007)&0.96 (0.004) \\
 \hline
    \end {tabular}
     \label{tab:graph}
     $^*$ scaled to be located between 0 and 1.
     \normalsize
\end{table}

\begin{figure}[h!]
{
\centering
\vspace{-10pt}
\includegraphics[scale=0.7]{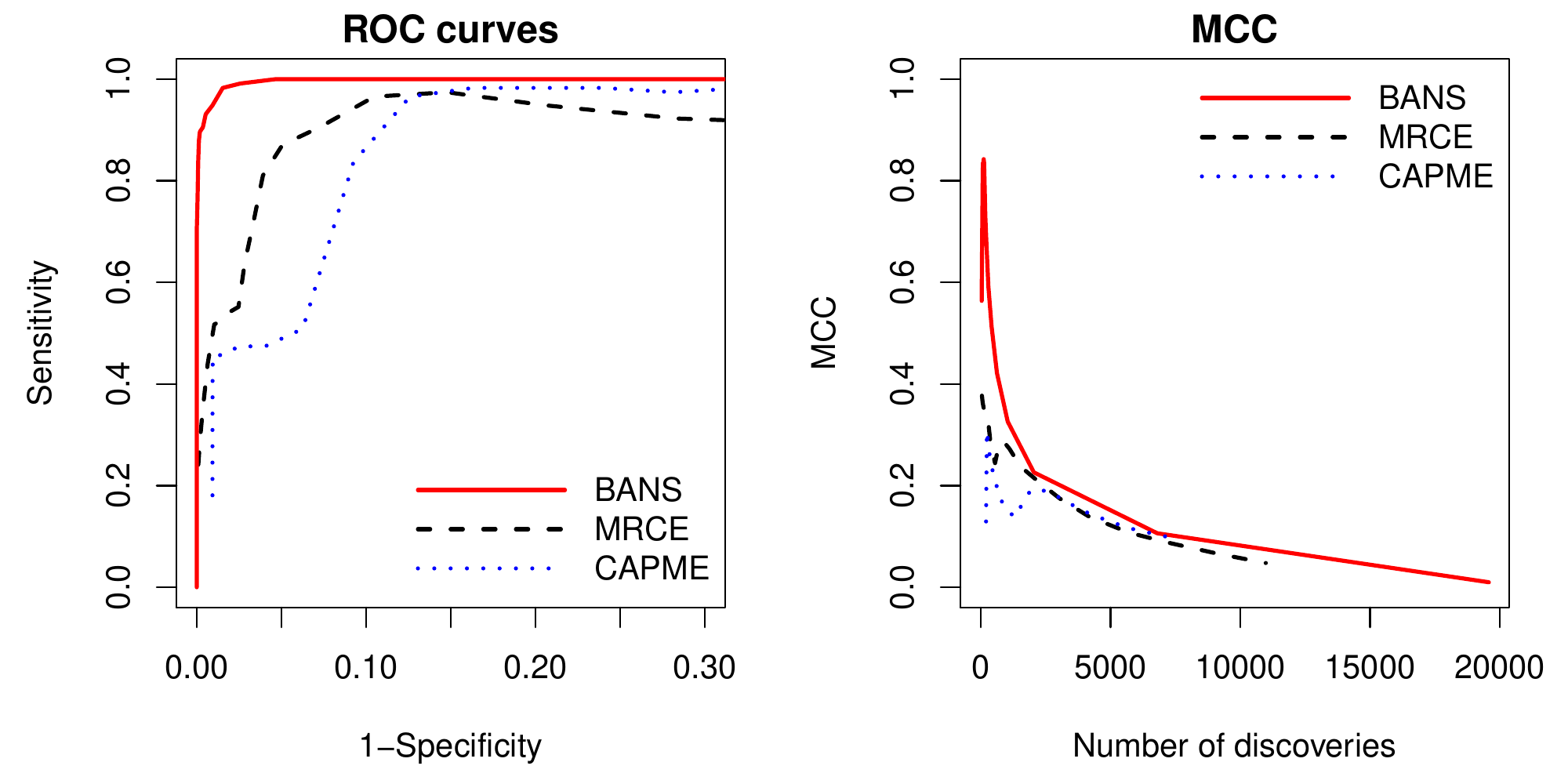}
\vspace{-20pt}
\caption{ROC curves and MCC curves for graph structure learning for $(p,n,q,p_E)$ = (200,100,10,0.03).}
\vspace{-20pt}
\label{fig:p200q10}
}
\end{figure}

To evaluate the sensitivity, specificity, MCC, and number of discoveries, we determine the tuning parameters of glasso, MRCE, CLIME and CAPME using five-fold cross-validation, and control the FDR at 0.1 for BANS. To further compare the performance on graph structure recovery, we obtain the receiver operating characteristic (ROC) curve and the MCC curve as the function of the number of discoveries for each simulation dataset by varying the tuning parameters for MRCE and CAPME, as well as the cutoff for the posterior marginal probability of edge inclusion for BANS. The sensitivity, specificity, MCC, and number of discoveries for the selected tuning parameters, and pAUC and AUC obtained from the ROC curves, are shown in Table~\ref{tab:graph}. In terms of graph structure recovery, our method yields better performance for all settings. The number of discoveries is very close to the number of edges in the true graph, while other methods tend to provide a much greater number of edges. The small standard errors for the number of discoveries suggest that our procedure is stable across simulation replicates. Figure~\ref{fig:p200q10} and Figure S10-S12 display the ROC curves and MCC curves averaged over 50 replications, which demonstrate that our method performs better than MRCE and CAPME. Our method shows better ROC curves and MCC curves for all four simulation settings across different tuning parameters. In the high-dimensional case, when $p=200$ and $n=100$ (Figure~\ref{fig:p200q10}), the sensitivities of MRCE and CAPME are flattened for some intervals of 1-specificity, and MRCE shows unstable variable selection performance, because the curves show a decreasing pattern as the tuning parameters decrease.
\vspace{10pt}
The Section S7 includes other extensive simulation studies. We evaluated the performance of BANS in simulation settings where data were generated from non-Gaussian, and the graphical structures were the same as the estimated DNA damage response network for 309 TCGA lung squamous cell carcinoma. We performed comparisons to (1) BANS-parallel (Section S6), (2) XMRF \citep{wan2016xmrf}, that fits UGs to mixed data types, (3) Neighborhood selection of UGs using Horseshoe prior \citep{carvalho2010horseshoe} and (4) objective Bayes fractional Bayes factor (OBFBF) \citep{SJOS:SJOS12273} using model averaging. Given the simulation settings where the data are generated from Poisson distribution, BANS showed better performance than XMRF (Section S7.2). For the evaluation of our prior choice, the point mass prior in equation (\ref{eq:finalprior}) showed superior performance than Horseshoe prior and OBFBF in estimating UGs (Section S7.3 and Section S7.4).  While BANS-parallel without symmetric constraint in the MCMC sampling showed smaller AUC and MCC values  than BANS, it still performed better than other methods such as MRCE and CAPME (Figure S8). Considering the gain in computational efficiency (Section S7.5), BANS-parallel is potentially useful alternative in high-dimensional setting. 
\subsubsection{Posterior inference on the signs}
The main focus of this section is to make inference on the signs of the edges, conditioned on the estimated undirected and directed structures using our node-wise regression approach. We define the sign for an undirected edge as the sign for its corresponding partial correlation. The partial correlation for an edge $v-w$ is $-\kappa_{vw}/\sqrt{\kappa_{vv}\kappa_{ww}}$, for which the sign is the same as that of the regression coefficient $\alpha_{vw}$ or $\alpha_{wv}$. The sign for a directed edge is straightforward, $sign(b_{vw})$ for $w\rightarrow v$. Thus, the signs of all estimated edges are obtained by structured estimation of the nonzero elements in $\mb{B}$ and $\bs{\mc{K}}$. The structured MCMC sampling scheme can be followed by the same procedure described in Section 5.1 and Section S3, given $\bs{\eta}$ and $\bs{\gamma}$. Then the posterior probabilities for negative or positive signs can be obtained for each edge.

\begin{figure}[h!]
{
\centering
\vspace{-20pt}
\includegraphics[scale=0.8]{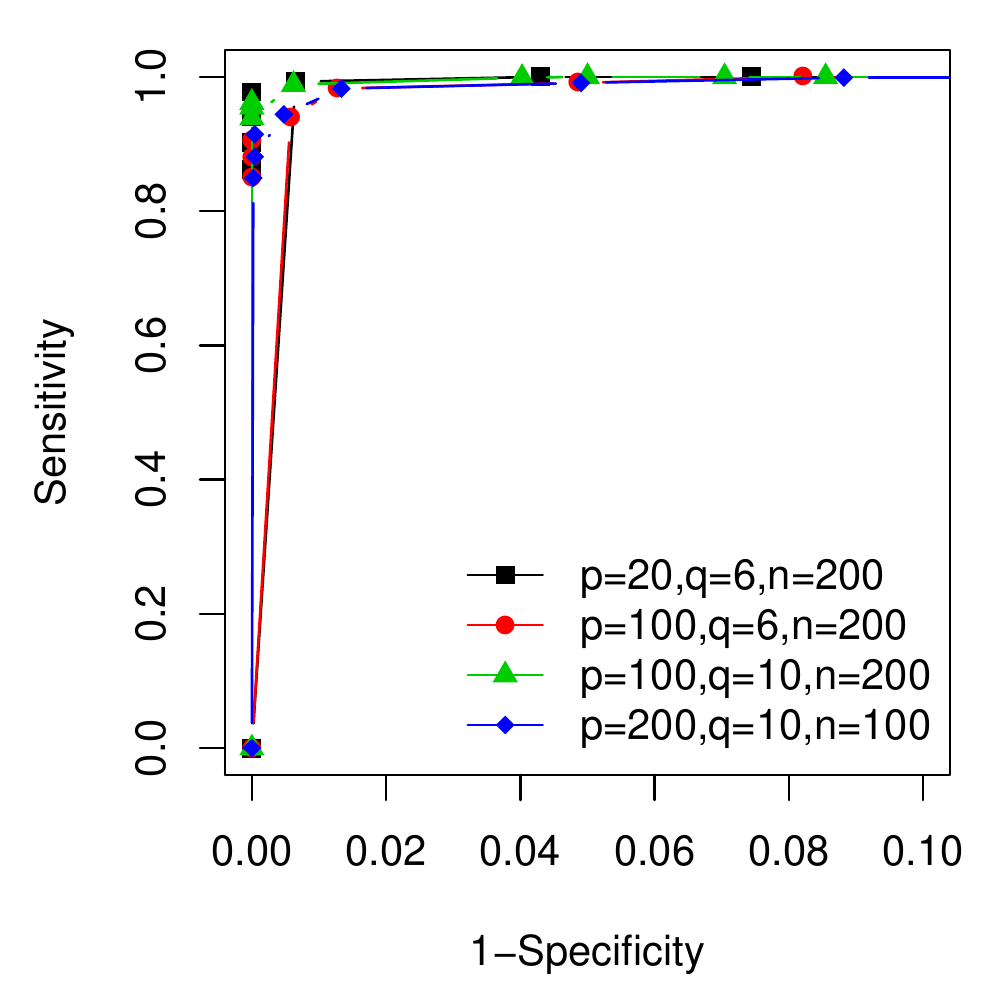}
\vspace{-20pt}
\caption{ROC curves for posterior inference on predicting positive signs from the structured estimation. The AUCs were 0.99 for all the scenarios.}
\label{fig:prop_matched}
}
\end{figure}
We investigate the performance of the inference on the sign using a simulation study. We declare that an undirected edge $v-w$ is positive, if $P(\alpha_{vw}>0|data)>\xi$, or negative otherwise; and a directe edge $w\rightarrow v$ is positive, if $P(b_{vw}>0|data)>\xi$, or negative otherwise. Figure~\ref{fig:prop_matched} shows ROC curves for predicting positive signs from the structured estimation, which was averaged over 50 replications. With AUCs 0.99 for all the four settings, we can conclude that signs are accurately estimated from our structured estimation given $\bs{\eta}$ and $\bs{\gamma}$, using our node-wise regression model.
\subsubsection{Sensitivity and Convergence}
We perform sensitivity analysis of our algorithm to priors (Section S4) and check the convergence (Section S5). In assessing the prior sensitivity of the model, we observe that the choice of $\lambda_\tau$ and $\delta_\tau$ while setting $c_{vw}^2 = 1/\lambda_\tau $ in equation (\ref{eq:finalprior}), which are the shape and scale parameters of the prior on the precision (inverse variance), $\kappa_{vv}$. Moreover, those hyperparameters contribute to the variance of the non-zero regression coefficients for undirected and directed edges. The average PPIs for non-zero edges was consistently higher (around 0.9) than those for zero edges (around 0.1) in the range from 1 to 10 of $\lambda_\tau$ and $\delta_\tau$ (Figure S2). For a convergence check, we observed that the trace plots from three independent chains showed that the number of edges included in the graphs had good mixing around a stable model size, and no strong trends (Figure S3).

\section{Pan-Cancer Network Anaysis of Multi-platform Omic Data}
\paragraph{Biological motivation} Crosstalk within signaling pathways and their perturbation by oncogenes limit single gene-based approaches to understanding cancer biology. Approaches have been developed for discovering mutations that perturb signaling networks (so-called network-attacking mutations) to understand how individual genomic variants initiate network perturbation \citep{creixell2012navigating,creixell2015kinome}. Although genomic variants are critical to understanding functional cancer networks, it is well-established that complex molecular networks and systems are formed by a large number of interactions of genes and their products, which operate in response to different cellular conditions \citep{bandyopadhyay2010rewiring,luscombe2004genomic}. Therefore, systematic approaches to unravelling cancer-specific rewiring of molecular networks are key to the successful identification of network-based drug targets for cancer treatment, in the paradigm of network medicine that acknowledges the application of network topology and dynamics towards identification of therapeutic targets \citep{califano2011rewiring,barabasi2011network}.

TCGA PanCancer Atlas initiative \citep{weinstein2013cancer} built a uniformly-processed dataset and a unified data analysis pipeline to develop an integrated picture of commonalities and differences across tumor types. Recent studies \citep{hoadley2018cell, sanchez2018oncogenic} reclassified human tumor types based on molecular similarity and investigated co-occurence of alterations in tumor signaling pathways, which differentiate between individual tumors and tumor types using TCGA pan-cancer data. \citep{akbani2014pan} investigated correlations between protein and other data types, such as mRNA, copy number, and mutation data across cancer. \citep{gong2017pancanqtl} performed expression quantitative trait locus (eQTL) analysis that focuses on the links between genotypes and gene expression for pan-cancer TCGA data. These methods only consider links between two platforms; they do not incorporate within platform dependencies. Using our BANS method, graph-based multi-level integration approach, our goal is to understand unified interplay within and between genomic, epigenomic, transcriptomic, and proteomic platforms to elucidate the commonalities and differences in systems across cancer types.

\begin{figure}[h!]
{
\centering
\vspace{-10pt}
\includegraphics[scale=0.4]{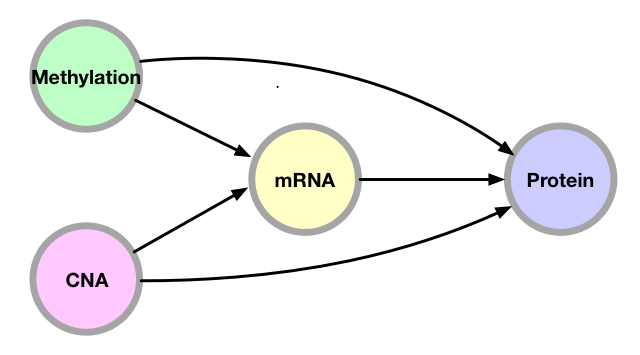}
\vspace{-20pt}
\caption{Inter-relationships between multi-platform data, copy number aberration (CNA), DNA methylation, nRNA expression, and protein.}\label{Fig:prior_ordering}
}
\end{figure}

\paragraph{Data structures} We applied our method to multi-omic datasets from the 7 TCGA tumor types, lung adenocarcinoma (LUAD, n=356), lung squamous cell carcinoma (LUSC, n=309), colon adenocarcinoma (COAD, n=338), rectum adenocarcinoma (READ, n=121), uterine corpus endometrial carcinoma (UCEC, n=393), ovarian serous cystadenocarcinoma (OV, n=227), and skin cutaneous melanoma (SKCM, n=333). For each of the tumor types, we included multi-platform data, DNA methylation, copy number alteration, mRNA expression data, and reverse phase protein array (RPPA)-based proteomic data. Genomic features from each data platform constitute a layer; the ordering of the layers follows biological justifications, because of the natural interplay among diverse genomic features: gene encoded by DNA is transcribed to mRNA, mRNA is translated to protein, and DNA methylation helps to regulate transcription  \citep{morris2017statistical} (Figure~\ref{Fig:prior_ordering}). 

Based on the principle that within-platform interactions arise from pathway-based dependencies that are altered across different tumor types, we selected 10 key signaling pathways, based on emerging literature on RPPA-based proteomic profiling of various tumor types \citep{akbani2014pan,cherniack2017integrated,li2017characterization}. The pathways and the gene membership for each pathway are listed in Table S1. We focus on building integrative networks for all combinations of the 10 pathways and the 7 cancer types. Using TCGA-Assembler \citep{zhu2014tcga}, we downloaded CNA, DNA methylation, mRNA expression, and protein expression data, then matched the samples across platforms. We applied BANS separately to each type of cancer and pathway combination (10$\times$7=70 analyses). The MCMC sampler was run for 10,000 iterations of burn-in, followed by 20,000 iterations as the basis for inference. For each analysis, we estimated the graph structure at FDR 0.1 and obtained the signs of each nonzero coefficient using the cutoff $\xi=0.5$ for the posterior probability of the signs from the structured estimation. Figures S13-S22 display the estimated networks for all pathways and cancer types.

\subsection{Global commonalities and differences in mlGGMs} 
\begin{figure}[h!]
{
\centering
\includegraphics[scale=0.35]{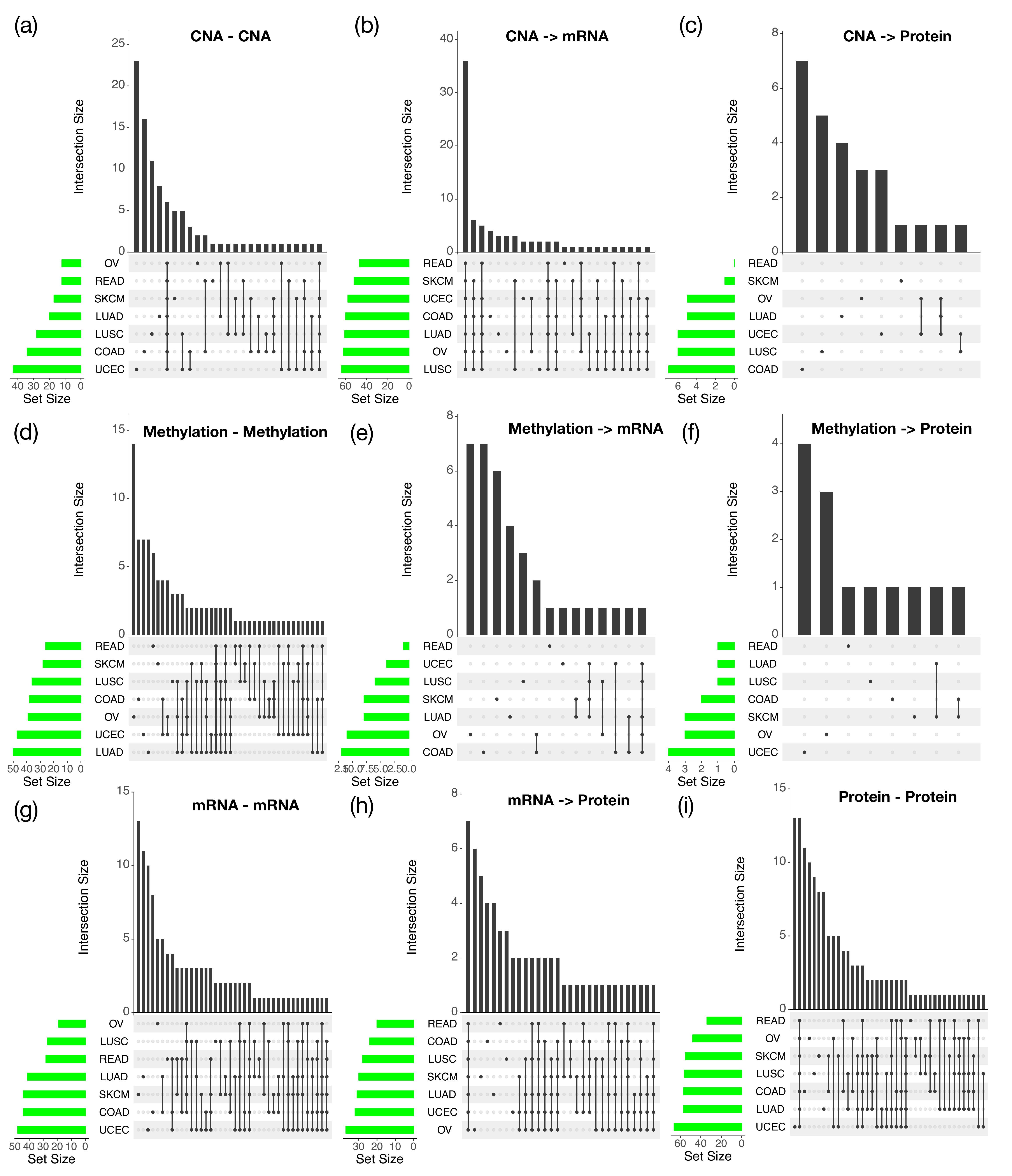}
\vspace{-30pt}
\caption{UpSet plots showing relationships of mlGGMs across all 10 pathways between 7 cancer types. Each column-wise bar corresponds to the number of exclusively intersecting edges that are shared by the cancer types represented by the dark circles, but not of the others, and each row-wise bar displays the total number of edges for the corresponding cancer type.}\label{Fig:UpSet}
}
\end{figure}

We first investigated the number of edges that are shared, differentiating across different cancer types for each of the possible edges at the platform level. For each combination of the cancer types and pathways, the mlGGM contains 4 layers at CNA, DNA methylation, mRNA expression, and protein expression, and allows 9 different types of dependencies, represented by the 4 types of undirected edges (i.e., CNA$-$CNA, methylation$-$methylation, mRNA$-$mRNA, and protein$-$protein) and the 5 types of directed edges (i.e., CNA$\rightarrow$mRNA, CNA$\rightarrow$protein, methylation$\rightarrow$ mRNA, methylation$\rightarrow$protein, and mRNA$\rightarrow$protein) (Figure~\ref{Fig:prior_ordering}). Across all pathways, our BANS method detected 433 (UCEC), 350 (SKCM), 328 (OV), 240 (READ), 391 (COAD), 394 (LUAD), and 361 (LUSC) directed/undirected edges in the estimated mlGGMs. Across all 10 pathways, we decomposed the edges by the aforementioned 9 types of dependences within and between the four layers, as well as the numbers of intersecting edges across the 7 cancer types are depicted using UpSet plots (shown in Figure~\ref{Fig:UpSet}).  UpSet plot is an effective visualization of intersections for more than three sets, and a more-scalable approach than Euler diagrams \citep{lex2014points}. For each of the 9 possible relations within and between layers, the UpSet plot contains column and row bar plots. The column bar plot encodes all edge set intersections in the columns of a matrix using a binary pattern, and renders bars above the matrix columns to represent the number of exclusively intersecting edges that are shared by the corresponding cancer types to the column, but not of the others. The row bar plot displays the total number of edges for each cancer type.

\noindent{\underline{Between platform regulatory relations:}} Transcriptional and translational effects represented by directed edges CNA$\rightarrow$mRNA (Figure~\ref{Fig:UpSet}-b) and mRNA$\rightarrow$ Protein (Figure~\ref{Fig:UpSet}-h) tend to be shared across cancer types, which is along expected lines of the basic biological mechanisms. For example, in Figure~\ref{Fig:UpSet}-b, 36 edges were shared by all 7 cancer types, and few unique edges to a cancer type were found. We found few regulatory edges from CNA to proteins, and from methylation to mRNA and Protein across cancer types (Figure~\ref{Fig:UpSet}-c,e,f). For example, for Methylation $\rightarrow$ mRNA in Figure~\ref{Fig:UpSet}-e, we found 7 edges unique to OV and COAD.

\noindent{\underline{Within platform conditional dependencies:}} The undirected edges represent dependencies within platform after taking out the regulatory effects from upstream platforms. The dependence structures within platforms tend to be unique to cancer type. For CNA (Figure~\ref{Fig:UpSet}-a), UCEC, COAD, LUSC, and LUAD had 23, 15, 11, and 7 edges that are not shared by any other cancer types, while 6 edges were shared across all cancer types. For Methylation (Figure~\ref{Fig:UpSet}-d), OV included 14 unique edges among 48 edges in total. For mRNA (Figure~\ref{Fig:UpSet}-g), SKCM had 13 of 45 edges that were unique. For protein (Figure~\ref{Fig:UpSet}-i), UCEC had the largest protein network, with 62 protein-protein interactions, 13 of which were unique to the cancer, while the same number of the edges were shared across all cancer types. 

\subsection{Pan-cancer network signaling} 
\begin{figure}[h!]
{
\centering
\includegraphics[scale=1.2]{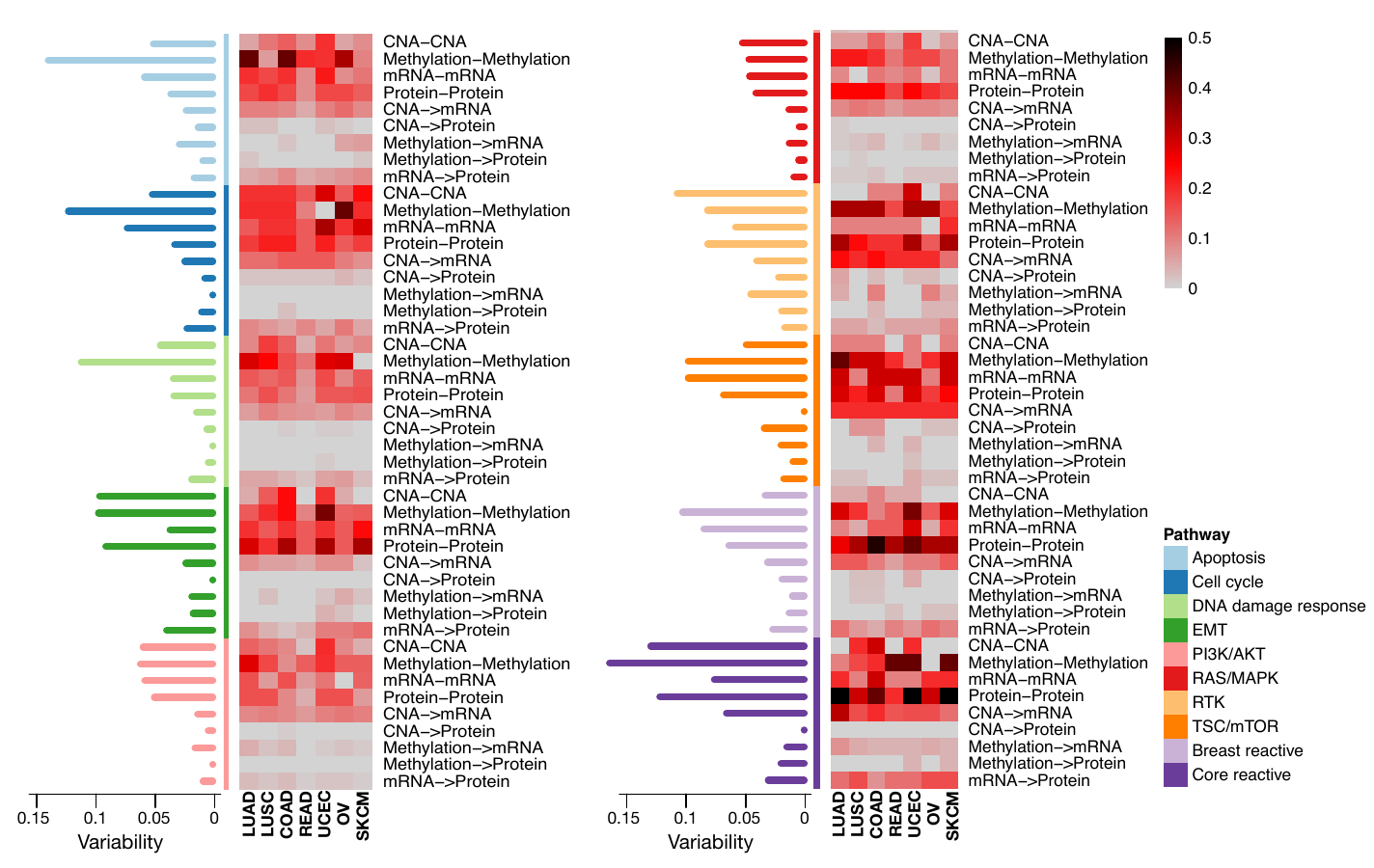}
\caption{Heatmap depicting connectivity score (CS) of the within- and between-layer sub-networks of the estimated mlGGMs across the 7 cancer types and 10 pathways. The scores are indicated on a low-to-high scale (grey-red-black). The standard deviations of the CS values across cancer types are displayed in the barplots.}\label{Fig:CSheatmap}
}
\end{figure}

We investigate the extent of cross-signaling within- and between- layers for each pathway across tumor types. For a given network, we define the ratio of the observed number of edges to the total number of possible edges as {\it connectivity score (CS)}: high (low) CS value indicates high (low) cross-signaling of the network. We also compute standard deviation of CS values across cancer types to represent that the levels of connectivity of the network are different across tumor types: the genes in the network are highly connected (high signaling) in some cancers, but have few connection (low signaling) in other cancers. Figure~\ref{Fig:CSheatmap} displays the CS values and the variability are displayed: CS ranged from 0 to 0.5, and the variability ranged from 0 to 0.17. The overall pattern of the heat map suggests that the within-layer sub-networks had a higher level of signaling than between-layer sub-networks. The core reactive pathway showed a high level of network signaling within platforms, and the highest level of protein-protein signaling across tumor types, which suggests that our estimation aligns well with the a {\it priori} functional characteristics of this pathway, which was defined on the basis of all tumor types \citep{akbani2014pan}. The CS was markedly different in methylation-methylation networks for most of the pathways, compared to other types of sub-networks: the apoptosis pathway for LUAD and COAD, cell cycle pathway for OV, EMT pathway for UCEC, TSC/mTOR pathway for LUAD, breast reactive pathway for UCEC, and core reactive pathway for READ, UCEC, and SKCM showed the high level of signaling for methylation-methylation. The transcriptional and translational effects that are represented by CNA$\rightarrow$mRNA and mRNA$\rightarrow$protein sub-networks showed the high-level of signaling across pathways and cancer types, compared to other between-layer networks, again following along expected lines of the biological hierarchy. 

\subsection{p53 Integrative Networks} In this section we focus on a gene/protein of particular interest (i.e., p53) across all tumor types. p53 is the most frequently mutated gene in human tumors, and has a central role as a tumor suppressor and novel therapeutic target \citep{bouchet2006p53,farnebo2010p53,levine2019targeting}. The transcription factor p53 is activated downstream of the DNA damage response (DDR) in reaction to cell stress, and mediates distinct outcomes of DDR signaling \citep{reinhardt2012p53}. The TP53 gene encodes the p53 protein that targets a large set of genes associated to apoptosis and cell cycle pathways \citep{bouchet2006p53,reinhardt2012p53}. Due to the high level of inter-connectivity of the DDR with other signaling networks, predicting the efficacy of treatment and designing an optimal combination therapy to target multiple genes will require a detailed understanding of the tumor-specific signals of other molecules.

\begin{figure}[h!]
{
\centering
\includegraphics[scale=0.28]{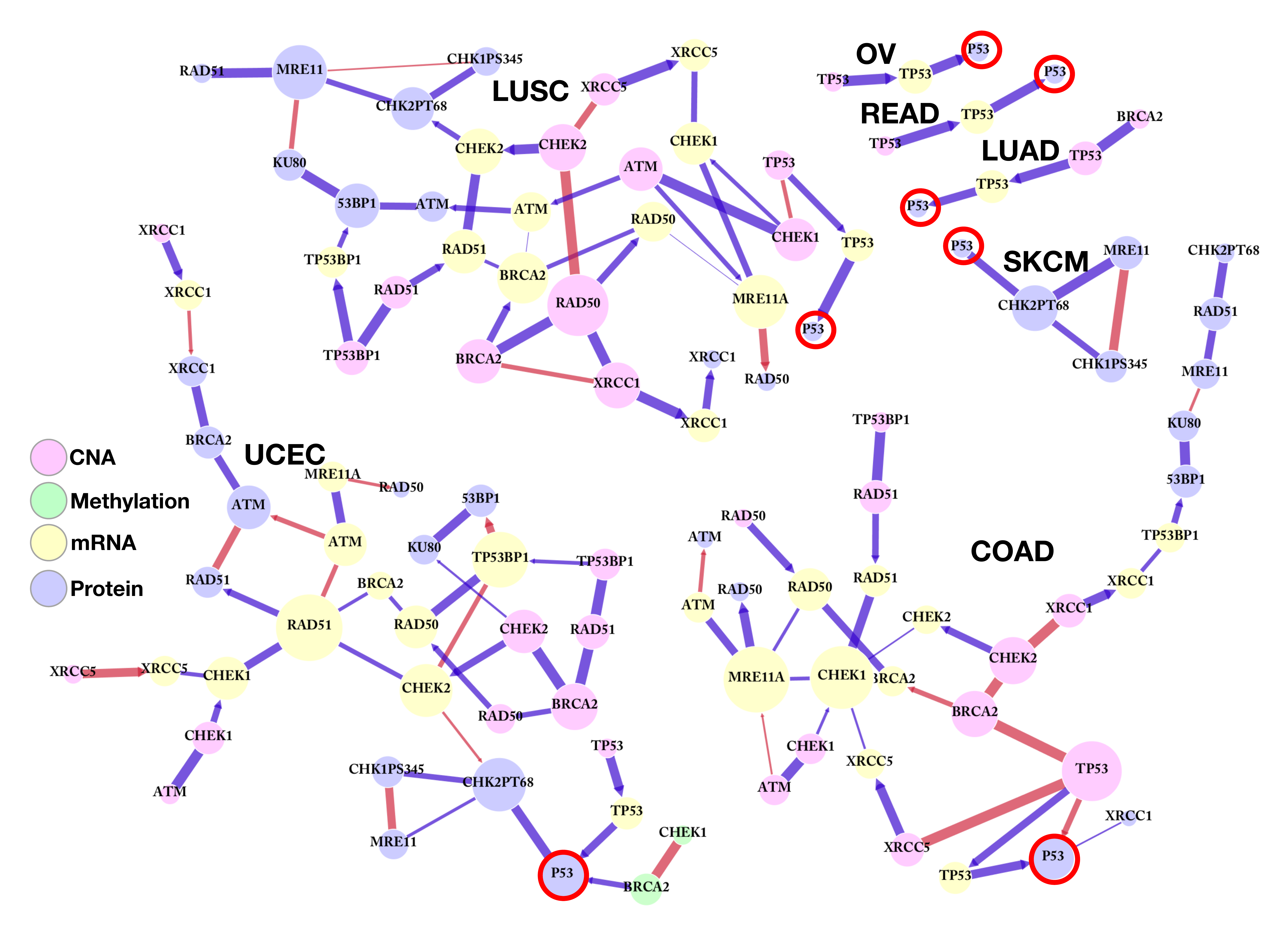}
\vspace{-30pt}
\hspace{-30pt}
\caption{Integrative sub-networks for the p53 protein (red circle), inferred from BANS for LUAD, LUSC, COAD, READ, UCEC, OV, and SKCM. Each connected component includes all nodes that are connected to the p53 protein by any lengths of paths, including both undirected and directed edges for each cancer. The colors of edges indicate the inferred signs of the edges: negative (red) and positive (blue). The sizes of nodes and the widths of edges are weighted by their degrees and posterior edge inclusion probabilities, respectively.}\label{Fig:net_TP53}
}
\end{figure}

Let $g_{ij}$ be the estimated posterior marginal probability of the edge ($i-j$ or $i\rightarrow j$) inclusion. We consider the estimated mlGGMs are weighted graphs, whose edges have been assigned the given posterior inclusion probabilities, and the degree of the nodes across all layers are defined as the summation of the weights for the edges that are connected to the node \citep{barrat2007architecture}: 
\begin{equation*}
W_{i}  = \sum_{j=1}^p I(i\rightarrow j \text{ or } i\leftarrow j \text{ or } i-j) g_{ij}, \forall i\in V .
\end{equation*}
The degree of the nodes in a weighted mlGGM measures the strength of nodes in terms of the total weight of their directed and undirected connections. With the goal of studying the underlying mechanism of p53 protein, we focused on the sub-networks of the estimated mlGGMs, which include the nodes connected to the p53 protein by any lengths of paths including both undirected and directed edges. Figure~\ref{Fig:net_TP53} displays the p53 integrative networks, where the edges are colored and weighted by the signs and the posteriors, $g_{ij}$, and the sizes of nodes are weighted by their degrees, $W_i$. 
\subsubsection{Translational findings}
Our results confirm that the transcriptional effect from TP53 at CNA to TP53 gene expression, and translational effects from TP53 gene expression to the p53 protein across 6 tumor types (all except for SKCM), all with positive regulations. In contrast, the SKCM network included no upstream regulatory effects (from DNA and RNA) and only included protein-protein interactions. UCEC and SKCM shared the same protein-protein interactions between p53 protein expression and T68 phosphorylated CHK2 (CHK2PT68): CHK2 is a protein kinase that is activated in response to DNA damage and directly phosphorylated p53 on serine 20, which provides a mechanism for increased stability of p53 \citep{hirao2000dna} and has been suggested as an anticancer therapy target given its role as a tumor suppressor \citep{zannini2014chk2}. These findings prioritize UCEC and SKCM as potential cancer types for CHK2 targeting. We also found epigenetic effects in the p53 sub-networks: only UCEC had a positive regulatory effect from BRCA2 DNA methylation, which was correlated with CHEK1 DNA methylation. The physical and functional interactions between BRCA2 and TP53 have been reported and hypermethylation of the CpG island in the promotors of BRCA2 gene involve their inactivation and therefore a higher risk of developing a tumor including uterine cancer \citep{rajagopalan2010mapping,bosviel2012brca2}. Epigenetic drugs have been developed and the anticancer effects are often tested using genome-editing technologies such as CRISPR-Cas9 systems, called Epigenome editing, which provides advantages over direct gene knockout based on RNA interference \citep{kungulovski2016epigenome}. Moreover, successes of epigenetic drugs have been reported by the important roles in synergy with other anticancer therapies or in reversing acquired therapy resistance \citep{morel2019combining}. Our findings through this deeper investigation of underlying biological mechanisms of p53 networks across multiple molecular levels and cancer types have the potential to facilitate the development of novel therapeutic strategies, specifically gene-drug interactions for single and combination agents.

\section{Discussion}
We propose a unified Bayesian framework to model the layered architecture of networks from multiple omic data. We employ a multi-layered Gaussian graphical model (mlGGM) to investigate the conditional dependencies among the variables from multiple layers, while accounting for the order defined by biological hierarchies. The mlGGM is built upon a multivariate regression framework with mean and residual precision parameters, for which zero structures represent undirected and directed edges in the chain graph. Our fully probabilistic formulation coherently converts the complex mlGGM into more tractable, node-wise multiple regression models, wherein the zero structures of the regression coefficients encompass both the undirected and directed edges. Our edge-specific variable selection priors on node-wise regression models allow for flexible modeling of any type of graph, without restriction to decomposable graphs, as well as the incorporation of relevant prior knowledge, while maintaining computational efficiency. We applied our Bayesian node-wise selection (BANS) method to Pan-cancer integrative network analysis and found structural commonalities and differences across cancer types. We also identified underlying mechanisms of the p53 protein, which is a novel drug target for cancer treatment. 

For identifiability, the main assumption in this article is that the layers have natural ordering. \cite{ma2008structural} and \cite{ha2015penpc} proposed methods to estimate a Markov equivalence class of a chain graph and a DAG (as a special case of the chain graph model) by recovering skeletons on its subgraphs with no ordering information. In the absence of natural ordering, a chain graph is not identifiable and only its Markov equivalence class can be identified. In many applications, the vertices are naturally partitioned into multiple ordered layers, along with time points or the intrinsic biological mechanisms.  

For chain graph models, there are two main types of conditional independencies implied by the LWF Markov property and the alternative Markov property (AMP). Through zero structures of the mean parameter $\mb{B}$ and the residual precision matrix $\bs{\mc{K}}$ in the mlGGM in (\ref{eq:1}), we describe the AMP on chain graph models. For the LWF Markov property in high-dimensional Gaussian settings, \cite{sohn2012joint,mccarter2014sparse} made a convex formulation by using a conditional joint distribution given the predictors. The AMP Markov property is coherent, however, with data generation (as described in Section S1) by the direct link between the zero structure of the parameters in the multivariate linear regressions, and the presence/absence of edges in the graph. 

For the resulting asymmetry of the structure, we impose the constraint on $\bs{\mc{K}}$ during MCMC. Although most graphical model selection approaches based on node-wise regressions focus on estimating the structure, we showed the numerical performance of estimating signs of the directed edges and undirected edges, based on the structured estimation from the conditional posterior $P(\bs{\alpha},\bs{\kappa},\mb{B}|\mb{Y}, \bs{\eta},\bs{\gamma})$, where $\bs{\eta}$ and $\bs{\gamma}$ are subjected to the structure $G$. 

Our node-wise regression-based formulation, BANS to jointly estimate the undirected and directed edges to a node provides a flexible modeling framework. In particular, the proposed approach can be extended to allow nonlinear regulatory relations between layers, instead of the linearity assumption on the parameter $\mb{B}$, and to infer multiple mlGGMs where some of the graphs may be unrelated, while others share common edges. BANS R codes implementing our method are available on \url{https://github.com/***/BANS}.








\clearpage

\bibliographystyle{Chicago}
\bibliography{ref_chaingraph}
\end{document}